
%
%
%
%
%
%
%

\input phyzzx

\sequentialequations

\overfullrule=0pt
\catcode`\@=11

\def \F{\phi}

\def \T{\theta}

\def \D{{\Delta}}
\def \DR{{\Delta}_{R}}

\def\NP{{\it Nucl. Phys.\ }}

\def\PL{{\it Phys. Lett.\ }}
\def\PR{{\it Phys. Rev.\ }}
\def\PRL{{\it Phys. Rev. Lett.\ }}

\def\JTP{{\it JETP \ }}
\def\JP{{\it J. Phys.\ }}

\def\Mod{{\it Mod. Phys. Lett.\ }}

\def\eqaligntwo#1{\null\,\vcenter{\openup\jot\m@th
\ialign{\strut\hfil
$\displaystyle{##}$&$\displaystyle{{}##}$&$\displaystyle{{}##}$\hfil
\crcr#1\crcr}}\,}
\catcode`\@=12

\REF\bkta{J. M. Kosterlitz and D. Thouless, \JP {\bf C6} (1973) 1181.}
\REF\bktb{V. L. Berezinskii, \JTP {\bf 34} (1972) 610.}
\REF\bktc{J. M. Kosterlitz, \JP {\bf C7} (1974) 1046.}
\REF\bal{S. Balibar and B. Castaing, Surf. Sci. Rep. {\bf 5} (1985) 87.}
\REF\lush{M. Lusher, G. Munster, and P. Weisz, \NP {\bf B180} (1981) 1;
C. Itzykson, M. E. Peskin, and J. B. Zuber, \PL {\bf B95} (1980) 259;
A. Hasenfratz, E. Hasenfratz, and P. Hasenfratz, \NP {\bf B180} 353.}
\REF\ian{B. Sathiapalan, \PR {\bf D35} (1987) 3277; I. Kogan, JETP Lett.
{\bf 45} (1987) 709.}
\REF\dav{F. David, \NP {\bf B257} (1985) 45; V. Kazakov, I. Kostov, and
A. Migdal, \PL {\bf B157} (1985) 295; J. Ambjorn, B. Durhuus, and
J. Frohlich, \NP {\bf B257} (1985) 433.}
\REF\kaza{V. Kazakov, \PL {\bf A119} (1986) 140; V. Kazakov and D. Boulatov,
\PL {\bf B186} (1987) 379.}
\REF\moo{G. Moore, hepth/9203061, YCTP-P1-92.}
\REF\hsu{D. Kutasov and E. Hsu, \NP {\bf B396} (1993) 93.}
\REF\mea{S. Dalley, \Mod {\bf A7}, 1651 (1992)}
\REF\wein{D. Weingarten, \PL {\bf 90} (1980) 280.}
\REF\stan{S. Leibler and L.Peliti, Phys. Rep. {\bf C184} (1989) 101 and
references therein.}
\REF\mig{D.~J.~Gross and A.~A.~Migdal, \PRL {\bf 64} (1990) 717;
M. Douglas and S.~Shenker, \NP {\bf B335} (1990) 635;
E.~Br\'ezin and V.~Kazakov, \PL {\bf 236B} (1990) 144.}
\REF\GKa{D. J. Gross and I. R. Klebanov, \NP {\bf B344} (1990) 475;
\NP {\bf B354} (1991) 459.}
\REF\kpz{V. Knizhnik, A. Polyakov, and A. Zamolodchikov, \Mod {\bf A3} (1988)
819; F. David, \Mod {\bf A3} (1988) 1651; J. Distler and H. Kawai, \NP {\bf
B321} (1989) 509.}
\REF\kazc{V. Kazakov and A. Migdal, \NP {\bf B311} (1989) 171.}
\REF\ivana{I. Kostov, \PL {\bf 297} (1992) 74. }
\REF\ivanb{I. Kostov, \NP {\bf B376} (1992) 539 and references therein.}
\REF\gin{P. Ginsparg, Los Alamos preprint LA-UR-91-999-mc.}
\REF\bipz{E. Br\'ezin, C. Itzykson, G. Parisi, and J-B. Zuber, Comm. Math.
Phys. {\bf 59} (1978) 35.}
\REF\gau{M. Gaudin, unpublished.}
\REF\mus{N. Muskhelishvili, {\em Singular Integral Equations} (Nordoff NV,
Groningen, Netherlands, 1953).}
\REF\mat{G. Moore, N. Seiberg, and M. Staudacher, \NP {\bf B362} (1991) 665.}
\REF\mec{S. Dalley, unpublished (to appear).}
\REF\desj{C. F. Baillie and D. A. Johnston, \PL {\bf B291} (1992) 233.}
\REF\simon{S. M. Catterall, J. B. Kogut, and R. L. Renken, \NP {\bf
B408}
(1993) 427.}

\def\eqaligntwo#1{\null\,\vcenter{\openup\jot\m@th
\ialign{\strut\hfil
$\displaystyle{##}$&$\displaystyle{{}##}$&$\displaystyle{{}##}$\hfil
\crcr#1\crcr}}\,}
\catcode`\@=12

\def\b{\beta}
\def\a{\alpha}
\def\l{\lambda}

\def\half{{1\over 2}}
\def\d{\delta}

\def\cm{{\tilde{m}}}
\def\r{\rho}
\def\DRT{{\tilde{\Delta}}_{R}}
\def\MT{\tilde{M}}

\nopagenumbers

{\baselineskip=12pt
\line{\hfil PUPT-1417}
\line{\hfil Revised January 1994}
\line{\hfil hepth/9309068}
}
\title{The Kosterlitz-Thouless Phenomenon on a Fluid Random Surface}
\author{Simon Dalley\foot{Present address: Department of Physics,
Theoretical Physics, Oxford University, Oxford OX1 3NP, United Kingdom.}}
\JHL
\abstract
The problem of a periodic scalar field on a two-dimensional dynamical
random lattice is studied with the inclusion of vortices in the
action.
Using a random matrix formulation, in the continuum limit for genus
zero
surfaces the partition function is found exactly, as a function of the
chemical potential for vortices of unit winding number, at a specific
radius in the plasma phase. This solution is used to describe the
Kosterlitz-Thouless phenomenon in the presence of 2D quantum gravity
as
one passes from the ultra-violet to the infra-red.
\endpage

\pagenumbers
\vsize=8.9in
\hsize=6.5in
\centerline{\caps 1. Introduction}
\bigskip

The Kosterlitz-Thouless (KT) phenomenon [\bkta, \bktb, \bktc]
has had far reaching implications
for our understanding of effectively two-dimensional (2D) physical systems.
Its myriad applications include those to 2D superfluids [\bkta],
crystal growth [\bal],
lattice gauge theories [\lush], and superstring theories [\ian].
While this phenomenon is usually studied on a fixed regular lattice,
it is known
that many aspects of 2D statistical mechanics become more tractable when
studied on lattices for which local
connectivity is itself a dynamical variable,
principally because they may be mapped to exactly solvable problems of
random matrices [\dav].
In the language of two-dimensional quantum field theory this corresponds
to the introduction of
2D quantum gravity.
They seem to display the same
critical phenomena as on fixed regular lattices but with mild modification of
critical exponents; for example, the Ising model transition is third order
rather than second [\kaza]. The possibility of finding some non-perturabtive
results for the KT vortex plasma motivates a study of this  phenomenon on such
fluid random surfaces. In this paper, using the example of a circular
scalar field
at a particular radius, it is shown by an exact  solution in the
continuum
limit how the gas of
unbound vortices leads to
topological order in two dimensions, making no assumptions about diluteness
save the restriction to vortex winding number one (the most relevant).

The random matrix model employed is a particular case of those studied
in [\mea] which may be reformulated as an $O(2)$ model on a dynamical
triangulation [\ivana, \gin]. On a single oriented link with periodic
boundary conditions, representing the circular target space, one has
an
$N$x$N$ complex matrix $M$ --- the models are the non-critical string
analogue of Weingarten's matrix model [\wein] ---
the full partition function being
$${Z = \int DM\ {\rm exp}\left( -\Tr [M^{\dagger}M] + {\kappa \over N}
\Tr [M^{\dagger}MM^{\dagger}M + MMM^{\dagger}M^{\dagger}] + h\sqrt{N}
\Tr [M+M^{\dagger}] \right)\ .}
\eqn\part$$
Expansion in $\kappa$, the bare cosmological constant, generates a
dynamical
quadrangulation (square simplices glued pairwise along edges) with the
edges embedded on the periodic link; also included in the action is a
coupling $h$ to vortices (and anti-vortices) of unit winding number,
which
cut small holes in the random surface that wind around the target
circle.
Gross and Klebanov have shown [\GKa]  that, in the double scaling limit
[\mig], the vortex-free $c=1$ matrix model partition function defined
on
a lattice is equivalent to that on the continuous real line provided
the
lattice spacing $a$ is less than some known critical value $a_{c}$. At
$a=a_{c}$ a phase transition occurs (believed to be of KT type) due to
the appearance of a marginal operator. The same holds true of the
complex
matrix models of ref.[\mea], from which one may deduce that \part\
represents the  case $a=a_{c}$; this means that the periodic link has
radius $r= a_{c}/2\pi = \half r_{sd}$, where $r_{sd}$ is the self-dual
radius. This places one deep in the purported vortex-plasma phase
since
unit charge vortices become relevant for $r< 2r_{sd}$.

It is precisely at $r=\half r_{sd}$ that one may reformulate the
problem
in terms of a solvable $O(2)$ model by introducing an inducing field
[\gin, \ivana]. Using an $N$x$N$ Hermitian matrix variable $\F$ one
can
rewrite \part\ as
$${Z \propto \int DM\ D\F\ {\rm exp} \left( -\Tr [M^{\dagger}M] -\Tr {\F}^2 +
\sqrt{2\kappa \over N} \Tr [M^{\dagger}M\F + MM^{\dagger}\F ]
+ h\sqrt{N}\Tr [M+M^{\dagger}]\right)\ ,}\eqn\partb$$
which after diagonalising $\F$ and performing the Gaussian
$M$-integrals
gives
$$\eqalign{Z(h,b_{0}) = \int \prod_{i=1}^{N} d\lambda_{i}\ {\rm exp}
 ( &-N\sum_{i} 2\lambda_{i}^{2} + \sum_{i\neq j} \log{(\lambda_{i}
-\lambda_{j})}  -\sum_{i,j} \log{(2b_{0} - \lambda_{i} -\lambda_{j})} \cr
& +4Nh^2 \sum_{i} (b_{0} - \lambda_{i})^{-1})\ .\cr}\eqn\gas$$
In eq.\gas\ $\lambda$ are the eigenvalues of $\F$, and there has been a
redefinition  $\sqrt{2\kappa} = 1/4b_{0}$ and rescaling $\lambda
 \to 2\lambda \sqrt{N}$ and $h^2 \to 8h^2/b_{0}$.
The case $h=0$ has been studied extensively by I.Kostov and
collaborators
[\ivanb] as a loop gas problem, who found $c=1$ critical exponents.
Thus
despite the presence of a marginal operator,
at the radius chosen, over
which one has little control, one can still study the KT phenomenon as
$h$ is turned on since the ultra-violet (UV) limit, corresponding to
$h=0$ as we shall see shortly, is that of a
massless
scalar theory.

At $h \neq 0$, the vortex--anti-vortex pairs
appear as an extra singular term in the potential
for this eigenvalue problem, which at large $N$ may be treated by saddle-point
methods [\bipz].
Indeed, the operator coupling to $h$ was identified already in the
loop gas approach [\ivanb] and has a clear geometrical meaning. If one expands
$Z$ \partb\ in Feynman diagrams, a typical piece of planar diagram will
appear as in Fig.1. A ``bug'' crawling on the surface goes $2\pi$ around the
circle every time it crosses an arrowed propagator $<M^{\dagger}M>$. Closed
loops of these (the loop gas) delineate regions of constant scalar
field, $X$ say, which since
the target lattice has only one link, takes only one value.
Vortex--anti-vortex
pairs appear connected by cuts in the $X$ field. If the length of a cut is
$l$, then in terms of the $\F$ field it acts like a hole of length $2l$ on
the surface with one marked point on the boundary. The other marked point,
given by the other vortex, is not independent since the two are constrained
to be length $l$ apart.
This is the origin of the macroscopic loop operator
coupling to
$h^2$ in \gas\ .

Thus the Boltzmann weights on a globally connected planar lattice take the
form
$$\log{Z}(b_{0},h)
= \sum_{\rm graphs} h^{\# {\rm vortices}} b_{0}^{-\rm total \
length\ of\ loops\ \& \ cuts}\ .\eqn\bol$$
One may therefore think of the arrowed lines as flux tubes which, because
of the peculiar single valuedness of the target lattice, carry a unit
charge's worth of electric flux for a Coulomb field. $-\log{h}$ is the
chemical potential for vortices, with small $h$ corresponding to the dilute
regime. Flux tubes densely populate the surface and the continuum limit for
area is achieved when they become infinite in length as $b_{0}$ is
tuned to its critical value [\ivanb].
\bigskip
\centerline{\caps 2. Saddle-Point Solution.}
\bigskip

Let us now study the effect of turning on $h$ by finding the
continuum limit of \gas\ .
The leading order of the large $N$ limit of $Z(h,b_{0})$ describes surfaces
of spherical topology. There exists a systematic approach to computing the
$1/N$ corrections [\ivana],
corresponding to higher genus surfaces, but here we will
concentrate on genus zero. Introducing an eigenvalue density
$\rho (\l)$
for the problem \gas\ , the saddle-point equation for $\r$ is
$$\int_{a}^{b} d\mu\ \r (\mu) \left( {\rm P} {1\over \l -\mu} + {1\over
2b_{0} - \l -\mu} \right) = 2\lambda - {2h^2 \over (b_{0} -\l )^2 }\
,\eqn\eqm$$
where the support of $\r$ is on the interval $[a,b]$. This integral equation
has been solved for $h=0$ by Gaudin [\gau].
The same method will be used to solve
$h \neq 0$ also. To transform \eqm\ to an integral equation with Cauchy
kernel make the following redefinitions
$$\l \to b_{0} - \sqrt{A +B\l}\ , A+B=(b_{0} -a)^2 \ ,A-B= (b_{0}-b)^2
\ .\eqn\red$$
Then
$$\int_{-1}^{1} d\mu\ \r (\mu)  {\rm P} {1\over \l -\mu}
= 2(\sqrt{A+B\l} -b_{0}) +  {2h^2 \over A+b\l }\ .\eqn\neqm$$
The inversion formula for this problem is [\mus]
$$\r (\l) = -{2\over \pi} \sqrt{1-{\l}^2 } \int_{-1}^{1} {d\mu \over
\sqrt{1-{\mu}^2 }} {\rm P} \left( {1\over \l -\mu} \right)
\left[ \sqrt{A + B\mu} - b_{0} + {h^2 \over A+B\mu} \right]\ .\eqn\inv$$
$A$ and $B$ are determined by the normalisation and positivity conditions
for $\r$;
$$\eqalign{ {B\over 2} \int_{-1}^{1} d\mu \ {\r (\mu)\over \sqrt{A+B\mu}}
&= 1 \cr
\int_{-1}^{1} d\mu {1\over \sqrt{1-{\mu}^2}} \left( \sqrt{A+B\mu} -b_{0}
+{h^2 \over A+B\mu} \right) & = 0\ .\cr}\eqn\cond$$

To study the critical behaviour one need only know the form of $\r$ \inv\
for $\l \to -1$, though it may actually be found completely in terms of
elliptic integrals. The continuum limit of surfaces is achieved as the
bare cosmological constant is tuned to its critical value, given by the
condition $b=b_{0}$, when the end of the support of $\r$ meets the
singularities in \eqm\ . After some algebra, and performing the $h$-term
integral, one can rewrite the solution \inv\ in the form
$$ \r (\l) = {4\over {\pi}^2} \sqrt{1-{\l}^2} \left( BK(k) -(A+B\l )\Pi
(\l ) -{\pi Bh^2 \over 2(A+B\l )\sqrt{A-B} }\right)\ ,\eqn\sol$$
where
$$\Pi (\l ) = \int_{0}^{\pi /2} {d\F \over \sqrt{1-k^2 \sin^{2}{\F}}}
{\rm P} \left( {1 \over \l -\cos{2\F}} \right)\ ,\eqn\elt$$
an elliptic integral of the third kind, while $k^2 = 2B/(A+B)$, $\mu =
\cos{2\F}$, and $K$ $(E)$ denotes the complete elliptic integral of the
1st (2nd) kind. From the positivity condition one finds
$$2E(k)\sqrt{A+B} -\pi b_{0} + {\pi B h^2 \over \sqrt{A^2 - B^2}} =0\
.\eqn\pos$$
Using \sol\ , the normalisation condition requires a little more manipulation
before it can be brought to the form
$$-{2\over {\pi}^2}(A+B)((1-k^2 )K^2 (k) -E^2 (k)) - {2h^2 \over \pi
\sqrt{A-B}} (K(k) + {\rm n.s.} ) =1\ .\eqn\nor$$
n.s. stands for terms non-singular in the limit $b \to b_{0}$ $(k \to 1)$,
which will be unimportant for the continuum limit. To approach it one
introduces the cut-off $\d \to 0$ and renormalised parameters $M$ and $\xi$
as follows [\ivanb]:
$$\eqalign{ {A\over B}-1 &=  M^2 {\d}^2 \cr
\sqrt{{A\over B} + \l} & = \xi \d\ .\cr}\eqn\rep$$
$M^{-1}$ is  the renormalised dieletric constant of the Coulomb gas,
$M$ the renormalised boundary cosmological constant coupling to the length of
cuts in $X$. Returning to \sol\ and evaluating it perturbatively in $\d$
one eventually finds $(\xi > M)$
$$\r (\xi , M) = -{16 \sqrt{A} \over {\pi}^{2}} \d (\log{M\d}) \sqrt{\xi^2
-M^2} - {h^2 \over 2\pi {\d}^2 } {\sqrt{\xi^2 -M^2} \over \xi^2 M} +
\ldots \eqn\ssol$$
where $\ldots$ stands for terms subleading as $\d \to 0$. The bare vortex
parameter $h$ couples to a relevant operator and therefore must be tuned
to zero for finite renormalised parameter. The scaling law follows from
\ssol\ by requiring the 2nd term to be of the same order as the first
$$ h^2 = - h_{R}^{2} {\d}^3 \log{M\d}\ .\eqn\law$$
This agrees with KPZ scaling [\kpz] at $r=\half r_{sd}$ but
there is a logarithmic scaling violation; such violations are well-known at
$c=1$, occurring also for the cosmological constant [\kazc]. The latter is
determined in terms of $M$ by expanding \pos\ and \nor\ in $\d$;
$$\eqalign{ \pi b_{0} & = 2\sqrt{2A}
( 1 - M^2 {\d}^2 \log{M\d}) + {\pi h_{R}^{2}
 \over M\sqrt{2}}{\d}^2 \log{M\d} + \ldots \cr
1 & = {4A \over \pi^2 } (1- 2M^2 \d^2 \log^{2}{M\d}) -{\sqrt{2}h_{R}^{2}
\over A\pi M} \d^2 \log^2{M\d} + \ldots \cr}\eqn\pcond$$
which, omitting $+ \ldots$ now, determines $A=\pi^2 / 4$ and
$$\DR = (M^2 + {\sqrt{2} h_{R}^{2} \over \pi^2 M}) \log^{2}{M\d}
\ ,\eqn\cosm$$
where $b_{0} = \sqrt{2} + \d^2 \DR$ defines the renormalised cosmological
constant conjugate to renormalised area of the surface $A_{R}$.
$\d$ is thus the physical cutoff.

One can now eliminate $M$ in
$\r$ \ssol\ in terms of the quantities $\DR, h_{R}$. Apart from the
logarithms,
\cosm\ is cubic in $M$ and it is convenient to use the parametric solution
$$M = 2\sqrt{\Lambda \over 3} \cos{\pi - \T \over 3}\ ,\eqn\mmm$$
where
$$ \cos{\T} = {h_{R}^{2} \over \sqrt{2} \pi^2 } \left( {3\over \Lambda}
\right)^{3/2} \ ,\ \Lambda = {\DR \over \log^{2} {M\d}}\ .\eqn\subm$$
The equation \cosm\ is sketched in Fig.2. \mmm\ corresponds to the largest
positive root, this being the physical one such that $M \to \infty$ as
$\DR \to \infty$. \ssol\ and \mmm\ constitute the solution for the scaling
part of $\r$, from which physical quantities may be calculated in the
continuum limit.

The first thing to notice is that \cosm\ , which is the positivity
condition for $\r$, has no solution for $h_{R}^{2} > \sqrt{2}\pi^2
(\Lambda /3)^{3/2}$. In this case the singular term in the potential \eqm\ has
become too strong and there is no smooth large $N$ limit.
Vortices contribute
vacuum energy which can be negative $\sim {\rm e}^{{\D}_{R}^{c}A_{R}}$
and
so the partition function at fixed
cosmological constant  does not converge in area $A_{R}$ if $\DR <
{\D}_{R}^{c}$  ($h_{R}$ too large).

As an intermediate step to calculating the partition function of spherical
surfaces, it is convenient to first derive the macroscopic loop
expectation at genus zero i.e. the partition function for surfaces with
the disc topology.
$$ \tilde{\omega} (z) = \int_{0}^{\infty} dl\ \omega (l)
 {\rm e}^{-zl}\ ,\eqn\disc$$
where $\omega (l)$ is the renormalised partition function for discs of
renormalised perimeter length $l$ with one marked point on the boundary.
Using \rep\ and \ssol\
$$\eqalign{\tilde{\omega} (z)& =
\int_{M}^{\infty} {d\xi \over z+\xi } \r (\xi ,M) \cr
& = -\log{M\d}\left( {4\over \pi} \sqrt{z^2 - M^2} \left( \log{z + \sqrt{z^2 -
M^2} \over M}\right) \left[ \sqrt{2} - {h_{R}^{2} \over \pi^2 M z^2 } \right]
+ {4h_{R}^{2} \over \pi^3} \left( {\pi \over 2z^2} - {1\over zM} \right)
\right)\ .\cr}\eqn\loop$$
In deriving \loop\ individual terms in the leading order which are analytic
in $z$ have been omitted since they correspond to surfaces of infinitesimal
area bounded by finite perimeter [\mat]. \loop\ for $h_{R} =0$ was given in
[\ivanb].
As a useful check one may verify that $\tilde{\omega}$ correctly solves
the finite difference equation derived by Kostov directly from the continuum
limit of \eqm\ (see e.g. ref.[\ivana]). The partition function for spherical
surfaces may be derived from $\tilde{\omega}(0) = d \log{Z}/dh^2$,
$$\eqalign{ \tilde{\omega}(0) & = 2\sqrt{2} (\log{M\d}) \left( M + {h_{R}^{2}
\over 2\pi^2 M^{2}} \right) \cr
& = 2\sqrt{2\DR \over 3} \left( \cos{\pi -\T \over 3} + \half {\cos{\T} \over
\cos^{2}{\pi - \T \over 3}} \right)\ ,\cr}\eqn\dee$$
by integrating w.r.t. $h^2$ at constant $b_{0}$. The constant of integration
is $c \sim \DR^2$ by dimensions, up to logarithms.  Using \subm\ it is
simpler to integrate w.r.t. $\T$, $dh^2 = -(\Lambda /3)^{3/2} \sqrt{2} \pi^2
\sin{\T} d\T + \ldots $
$$\log{Z} \propto {\DR^2 \over \log^2{\d^2 \DR}} \left( \sin^{2}{\T -\pi \over
3} + {1\over 8} \log{\cos{\T - \pi \over 3}} -\half \sin^{4}{\T -\pi \over 3}
\right) + c\ .\eqn\sph$$

\noindent This represents the continuum limit where all subleading
terms $+ \ldots$
have been dropped. Equations \loop\ and \sph\ are the main results of this
paper, representing the continuum limit on genus zero surfaces for the
vortex gas with arbitrary coupling.

One can use these results to investigate the number of effective degrees
of freedom in various regimes (see refs.[\hsu, \moo] for a similar
non-perturbative study of the Sine-Gordan model coupled to 2D
gravity).
 In particular one can use \sph\ to find the
string susceptibility $\gamma_{\rm str}$, given by the fixed area partition
function $\log{Z(A_{R})} \sim A_{R}^{-3 + \gamma_{\rm str}} {\rm exp}(\DR^c
A_{R})$.
The UV limit $A_R \to 0$ occurs for $\DR \to \infty$, in which case \sph\
yields $\gamma_{\rm str} =0$, as appropriate for a massless scalar field. As
the scale $A_R$ is increased above that set by $h_{R}^{-4/3}$ the screening
effects of the vortex condensate come into play and the behaviour changes. To
determine $\gamma_{\rm str}$ more generally one must identify the effective
cosmological constant, given by the sum of contributions from gravity $(\DR)$
and matter $(-\DR^c )$, which is conjugate to area. This constant is zero
at the critical point of \cosm\ allowing identification of $\DR^c$. Since
$\T_{c} =0$ one finds
$$\eqalign{ M_{c}^{2} \log^{2}{\d M_{c}}& = {\DR^c \over 3} \cr
{h_{R}^{2} \over \sqrt{2} \pi^2 } \left( {3 \over (\DR^{c})^2 } \right)^{3/2}
\log^{3}{\d M_c} & = 1 \cr
\Rightarrow M_c & =  \left( {h_{R}^{2} \over \sqrt{2} \pi^2} \right)^{1/3}
\ .\cr}\eqn\mct$$
As usual, the presence of logarithms complicates $\DR^c$ which moreover
satifies a transcendental equation. $\DR^c = 3 M_{c}^{2}$ neglecting
 logarithms.

To find $\gamma_{\rm str}$ in the infra-red (IR)
limit $A_{R} \to \infty$ one can
proceed analytically. Defining $\DR = \DR^c + \DRT$ and $M = M_c + \MT$,
for small $\T$ one may expand all formulae for $\DRT \ll \DR^c , \MT \ll M_c$
and find expressions for the small variables; even keeping track of logarithms
is not too tedious. Expanding about \mct\ one finds
$$\eqalign{\MT = & {1\over \log{M_c \d}}
\left( {\DRT \over 6} \sqrt{3\over \DR^c}
+ {\T \over 3} \sqrt{\DR^c \over 3} + \ldots \right) \cr
-\half \T^2 + {\T^4 \over 4!} - \ldots = & -{3\over 2} \left( {3\over \DR^c}
\right)^2 M_{c}^{2} \DRT \log^{2}{\d M_c} + {2\MT \over M_{c} \log{M_c \d}}
+ \ldots \cr}\eqn\exa$$
and expanding $\DRT$ in $\T$ gives
$$\T = 3\sqrt{\DRT \over \DR^c} + O(\DRT^{3/2})\ .\eqn\thet$$
Expanding the partition function \sph\ in $\T$ one finds
$$\log{Z} = {(\DR^c )^2 \over \log^{2}{\d^2 \DR^c}} \left( \a + \b \T^2
+ \gamma \T^5 + \ldots +(\epsilon + \nu \DRT + \eta \DRT^2 + \ldots )\right)\
,\eqn\near$$
where $\a , \b ,\gamma ,\ldots$ are known constants and the constant of
integration $c= \epsilon +\ldots $ is analytic  as $\DRT \to 0$ since
it does not depend upon $h$, only $\DR$. Thus
$$\log{Z} = {\rm analytic} + {{\rm const.} \over \log^{2}{\d^2 \DR^c}}
{\DRT^{5/2} \over \sqrt{\DR^c}} + O(\DRT^{7/2})\ ,\eqn\far$$
confirming that $\gamma_{\rm str} = -1/2$ in the IR limit, where the cuts
in $X$ disorder it to leave pure gravity. Note that the
analytic terms as $\DRT \to 0$ come from the lower cutoff on the area
integral, most importantly they do not contain logarithms of $\DRT$.

The change in behaviour from the UV to the IR can be understood from the
Coulomb gas in terms of screening effects [\bkta]. At small areas $\DR \to
\infty$
and the renormalised $M^{-1} \to 0$ \cosm\ , all flux lines being of the order
of the UV cutoff on the surface; this is the $c=1$ regime where vortices are
strongly bound to anti-vortices. At large areas $\DR \to \DR^c$ and $M \to
M_{c}$ \mct\ . When $h_{R}=0$, since the surface is dense with flux loops,
to achieve large area the dielectric constant is tuned to $\infty$ ($M_c =0$).
In the presence of the charge gas ($h_{R}\neq 0$) the inverse permittivity
is $M-M_c$, the (non-universal) shift $M_c$ being due to dielectric
polarization. It clearly required the non-perturbative treatment to see
this since \mct\ is not analytic as $h_R \to 0$.
\bigskip
\centerline{\caps 3. Conclusions.}
\bigskip

Unfortunately it appears difficult to conceive of a general method for
obtaining
exact results for $r \neq \half r_{sd}$ or higher vortex winding
number.
However one may add the winding number $\pm 2$ operator, $\cm_2
\Tr [MM + M^{\dagger}M^{\dagger}]$, to \part\ and the integrals are
still
Gaussian, yielding
$$\eqalign{Z = \int \prod_{i} d\l_i \ {\rm exp} (& \sum_{i\neq j} \log{(\l_i
-\l_j)} - N\sum_{i} 2\l_{i}^{2} -\half \log{((2b_{0} -\l_{i} - \l_{j})^2
-(\cm_2)^2 )} \cr
& +8N\cm_1 \sum_{i} (b_{0} -\half \cm_2 -\l_i )^{-1})\ .\cr}\eqn\newz$$
instead of \gas\ , with $h= \cm_1$.
The geometrical meaning of the last two terms in the action \newz\ is
illustrated in terms of flux lines in Fig.3. The saddle-point equation to
be solved is
$$\int_{a}^{b} d\mu \ \r (\mu) \left( {\rm P} {1\over \l -\mu} +\half {1\over
2b_{0} -\cm_2 -\l -\mu } +\half {1\over 2b_{0} + \cm_2 -\l -\mu }\right)
= 2\l - {2({\cm_1})^2 \over (b_{0} -\half \cm_2 -\l )^2 }\eqn\newsp$$
but this is more difficult than \eqm\ ; the author has not managed to solve
it to the extent that useful information can be gleaned, for example on
questions of multicriticality as $\cm_2$ is tuned as a function of $\cm_1$.
One may however give a hand-wave argument that the IR limit
is pure gravity. Consider $\cm_1=0$ and $\cm_2 \neq 0$
in \newsp\ for simplicity.
According to the discussion of the previous section, the scaling regime
is described by $(\l'  +\half \cm^{R}_{2})\d  = b_{0} - \l $ and
$$\int_{b'}^{\infty} d\mu' \r (\mu') \left( {\rm P} {1\over \l' -\mu'} -
\half {1\over \l' + \mu'} - \half {1\over 2\cm^{R}_{2} + \l' + \mu'}\right)
= 2(\d \l' + \d \half \cm^{R}_{2} - b_0 )\ ,\eqn\sca$$
where ${\cm}^{R}_{2} =
\cm_2 /\d$ is consistent with KPZ scaling. The IR limit is
equivalent to $\cm^{R}_{2} \to \infty$ in which case the 3rd term in the
kernel
of \sca\ may be dropped, leaving the saddle point problem of the O(1) model
in Kostov's classification of the dense phase of $O(n)$ models
[\ivanb]
i.e., pure gravity.

For the more general models of ref.[\mea] one can ask about the extent
to which similar methods of solution may be applied, in particular the
introduction of induction fields to obtain plaquettes from an action
quadratic in $M$.
A particularly interesting problem concerns the effects of an
extrinsic curvature dependence, which can smooth badly behaved surfaces
[\stan]. This may be included in the
Weingarten-type [\wein]
models by introducing new flavours of link matrix. For example
at $c=2$, two flavours $M,N$ suffice since the worldsheet normals are $\pm 1$.
The extrinsic curvature, given by $\half (1-\cos{\T}) = 0,1$ for parallel
($\T = \pi$) or anti-parallel ($\T = 0$) neighboring plaquettes where $\T$
is the angle subtended, can be included by using an action
$$\sum_{l} \Tr [M^{\dagger}(l)M(l) + N^{\dagger}(l)N(l)] + K \Tr[N^{\dagger}
(l)M(l) + M^{\dagger}(l)N(l)] - \kappa \sum_{\rm plaq} \Tr[P_M + P_N]\
,\eqn\ext$$
where $\sum_l$ is the sum over all links of a square lattice;
$P_M$ is the product of $M$'s around  plaquettes of orientation $+$
and $P_N$ the product of $N$'s around  plaquettes of orientation $-$; $K$
is the bare extrinsic curvature coupling. The $c$-dimensional generalisation
has $Z_c$ symmetry and $c(c-1)$ flavours, and one assumes that the
continuous symmetry would be restored at any critical point. These models are
difficult to solve in full generality, but simplified
versions which still incorporate extrinsic curvature can be reduced to
eigenvalue problems. This will be dealt with in a future publication [\mec]

To summarize, in this paper the matrix model of random surfaces
introduced
 in ref.[\mea]
has been further investigated and applied to the two-dimensional vortex
plasma in the presence of gravity. The exact solution on genus zero was
found at half the self-dual radius as a function of renormalised
cosmological constant and vortex chemical potential \loop\ \sph\ . In
general these matrix models seem to give new possibilities for exact
solution where the traditional Hermitian models are too difficult, the
principle improvements being:
\item
{1.} Vortex operators are manifest and easily manipulated.
\item
{2.} Exact solutions are possible even in the presence of non-tree-like
embeddings.
\item
{3.} Extrinsic curvature can be trivially introduced and, in certain
cases [\mec], leads also to solvable models.
\item
{4.}  The matrix models have gauge symmetry [\mea] and are in fact
closely related to
conventional non-linear lattice gauge theories, for which there is an
abundance of expertise available.

The accompanying geometrical picture confirms the KT hypothesis of
unbound vortices, which in the renormalised scheme are joined by Coulomb
flux tubes, cuts in the periodic scalar field that disorder it at large
length scales. Hopefully this approach can be further developed to extend
the remarkable non-perturbative answers which matrix models furnish.
\bigskip

\ack
I have benefitted from discussions with I.Klebanov, I.Kostov,
M.Martin-Delgado,
and M.Staudacher.
Financial support was provided by  S.E.R.C.(U.K.) post-doctoral fellowship
RFO/B/91/9033.
\bigskip

Note Added: The reader's attention is also drawn to recent numerical
studies
of the XY model on random surfaces [\desj, \simon].
\endpage

\centerline{FIGURE CAPTIONS}
\bigskip

\item
{\bf Fig.1.} A typical piece of planar Feynman graph for action \part\ . A
single vortex--anti-vortex pair and three ``vacuum'' flux loops are shown
arrowed. Dotted line is $<\F \F >$ propogator.
\item
{\bf Fig.2.} A sketch of the cubic equation \cosm\ .
\item
{\bf Fig.3.} Vortex flux configurations corresponding to the expansion
of action \newz\ in powers of $\cm_1$ and $\cm_2$.
\endpage

\refout

\bye